\documentclass[letterpaper,preprintnumbers,twocolumn,eqsecnum,superscriptaddress,aps,nofootinbib]{revtex4}
\usepackage{amssymb}\usepackage[centertags]{amsmath}\usepackage{txfonts}\usepackage{epsfig}\usepackage{bm}
\usepackage{color}\usepackage{graphicx,graphics}\usepackage{multirow}\usepackage{float}\usepackage{ulem}
\usepackage{hyperref}\usepackage{setspace}\usepackage{slashed}\usepackage{booktabs}
\hypersetup{colorlinks=true, citecolor=blue, linkcolor=red, filecolor=black,urlcolor=blue}
\allowdisplaybreaks[4]

\begin{document}

\title{Neutrino-jet correlations in charged-current SIDIS}

\author{Weihua Yang}
\affiliation{College of Nuclear Equipment and Nuclear Engineering, Yantai University,\\ Yantai, Shandong 264005, China}

\author{Jing Zhao}\thanks{Contact author:zhaojingzj@sdu.edu.cn}
\affiliation{Key Laboratory of Particle Physics and Particle Irradiation (MOE), Institute of Frontier and Interdisciplinary Science, Shandong University, \\Qingdao, Shandong 266237, China}

\author{Zhe Zhang}\thanks{Contact author:zhangzhe@impcas.ac.cn}
\affiliation{Southern Center for Nuclear-Science Theory (SCNT), Institute of Modern Physics, Chinese Academy of Sciences, Huizhou 516000, China}


\begin{abstract}
Charged-current deep inelastic scattering plays a significant role in determining parton distribution functions with flavour separation.
In this work, we present a systematic calculation of the charged-current semi-inclusive deep inelastic scattering (SIDIS) in the $eN$ collinear frame up to twist-3 level at leading order. 
Semi-inclusive refers to the process in which a jet is detected in addition to the scattered neutrino. We focus on neutrino-jet correlations in our calculation. We first present the differential cross section in terms of structure functions, followed by the differential cross section expressed in term of transverse momentum dependent parton distribution functions. We derive the complete set of azimuthal asymmetries and intrinsic asymmetries. 
We also introduce an observable $A^C$, defined as the ratio of the difference to the sum of differential cross sections for electron and positron semi-inclusive deep inelastic scattering.
We notice that $A^C$ provides a sensitive probe for valence and sea quark distribution functions

\end{abstract}

\maketitle

\section{Introduction}

The lepton-nucleon deep inelastic scattering (DIS) has achieved great success in our understanding of the partonic structure of the nucleon.  It will still play an important role in the future Electron-Ion collider (EIC) \cite{Accardi:2012qut,AbdulKhalek:2021gbh,Anderle:2021wcy} experiment to explore the spin and three-dimensional structure of the nucleon over wide kinematic regions. Thanks to the factorization theorem \cite{Collins:1989gx}, the cross section of DIS can be factorized into two parts, one is the hard part which can be calculated with perturbative theory, the other is the soft part which involves parton distribution functions (PDFs).  The one-dimensional or collinear PDFs is studied in the inclusive DIS, where only the scattered lepton is detected with high precision.  However, inclusive DIS can not access three-dimensional or transverse momentum dependent PDFs (TMDs), since no transverse momentum scale is introduced. To solve this problem, one needs to consider the semi-inclusive DIS (SIDIS), where a hadron or a jet is also detected in addition to the scattered lepton. 

For the hadron production in SIDIS \cite{Mulders:1995dh,Bacchetta:2006tn}, fragmentation functions (FFs) which describe the formation of hadrons from a partonic initial state are involved. The cross section will be given in terms of convolutions of PDFs and FFs. Fragmentation functions are nonperturbative quantities and can only be determined by experimental measurements. 
Comparing to hadron production SIDIS,  the jet production one does have a simpler form of theoretical formulation and not introduce extra uncertainties from FFs. This is helpful to improve the measurement accuracy. Furthermore, the jet can be a direct probe of analyzing properties of the partonic structure of nucleon. For example, the transverse momentum of the jet is equal to that of the quark in the $VN$ collinear frame \cite{Yang:2022xwy,Yang:2023vyv,Yang:2023zod,Wu:2023omf}. Here $V$ denotes the intermediate exchange boson, whose transverse momentum  is defined as zero. For these reasons, jet production SIDIS for TMDs studies has attracted a lot of attentions. Nevertheless, jet production SIDIS process can not cover the low energy kinematic regions and access the chiral-odd TMDs due to the conservation of the helicity. The hadron production SIDIS  process therefore has to be considered to study chiral-odd TMDs.  We note here that proposals for exploring the chiral-odd TMDs via  jet fragmentation functions have been studied recently in Refs. \cite{Accardi:2017pmi,Accardi:2019luo,Accardi:2020iqn,Kang:2020xyq}.



In this paper, we calculate the charged-current jet production SIDIS process in the $eN$ collinear frame. Because charged-current interactions can be used as a clean probe to separate quark flavors, which cannot be achieved in the neutral-current DIS alone. Furthermore, charged-current interactions with the high statistics samples can offer precision measurements of electroweak parameters, e.g., weak mixing angle \cite{Boer:2011fh} which is used to determine the running of $\sin^2\theta_W$ as a function of $Q^2$.
In our analysis, the jet is simplified as a quark. Considering the conservation of momentum, the jet and the scattered electron are produced back-to-back in the plane perpendicular to the beam direction. However, the intrinsic transverse momentum of the quark or higher order gluon radiations would break the balance and induce measurable effects. We notice that these effects induced by the intrinsic transverse momentum of the quark provide a favourable palace to study twist-3 TMDs. To this end, we present systematic calculations of twist-3 measurable quantities for the neutrino-jet correlations ($j=l'+k'$) in the $eN$ collinear frame. Our method is consistent with that of Refs. \cite{Song:2010pf,Song:2013sja,Wei:2016far,Chen:2020ugq,Yang:2020qsk}, which focused on higher twist calculations at leading order. On the other hand, 
calculations can also be done from a higher-order perspective  \cite{Gutierrez-Reyes:2018qez,Gutierrez-Reyes:2019vbx,Liu:2018trl,Liu:2020dct,Kang:2020fka,Arratia:2020ssx,Arratia:2019vju,Arratia:2022oxd}. In this paper, we calculate both the electron and positron scattering off a nucleon (nucleus) and write theoretical expressions in a general form. We first present the differential cross section in terms of structure functions, followed by the differential cross section given in terms of TMDs at twist-3 level. We further calculate the complete azimuthal asymmetries and intrinsic asymmetries. We also introduce an observable $A^C$, defined as the ratio of the difference to the sum of differential cross sections for electron and positron SIDIS processes. We notice that $A^C$ is sensitive to quark distribution functions and can be used to study the violation of strange-antistrange symmetry. 
We note here that our consideration is different from that in Ref. \cite{Arratia:2022oxd}, in which both neutrino-jet correlations and hadron-in-jet spin asymmetries were calculated at leading twist. Specially, we focus on the neutrino-jet correlations and perform the calculation at twist-3 level.

To be explicit, we organize this paper as follows. In Sec. \ref{sec:generalcs}, we present the formalism and conventions used in this paper and give a brief introduction of the $eN$ collinear frame. In Sec. \ref{sec:structure}, we present a systematic calculation of the differential cross section of the jet production SIDIS in terms of structure functions. The systematic calculations in the parton model up to twist-3 level are shown in Sec. \ref{sec:partonm}. Measurable quantities and numerical estimates are given in Sec. \ref{sec:measurable}. A brief summary is given in Sec. \ref{sec:summary}.

\section{The formalism} \label{sec:generalcs}
In this paper, we consider charged-current semi-inclusive deep inelastic scattering (SIDIS), in which both a jet and the scattered lepton (neutrino or antineutrino) are detected in the final state. 
The electron scattering process is labeled as follows,
\begin{align}
 & e^-(l) + N(p,S) \rightarrow \nu_e(l^\prime) + jet(k^\prime) + X,
\end{align}
where $e^-$ denotes an electron with momentum $l$, $N$ represents a nucleon with momentum $p$, and the $jet$ corresponds to a quark with momentum $k^\prime$, which is observed as a jet of hadrons in experiments.
We require that the jet is in the current fragmentation region.
For the positron scattering process, $e^-$ is replaced by $e^+$, and the final-state neutrino $\nu_e$ is replaced by the antineutrino $\bar{\nu}_e$.
To be explicit, we list the standard variables for the jet production SIDIS,
\begin{align}
  & s=(p+l)^2, && x=\frac{Q^2}{2 p\cdot q}, &&  y=\frac{p\cdot q}{p \cdot l}, \label{f:sidisvar}
\end{align}
where $Q^2 = -q^2=-(l-l')^2$. 

Regardless of whether the process involves electron or positron scattering, the differential cross section can always be expressed as the contraction of the leptonic tensor and the hadronic tensor,
\begin{align}
  d\sigma = \frac{\alpha_{\rm em}^2}{2sQ^4}A_W L_{\mu\nu}(l, l^\prime)W^{\mu\nu}(q,p,S,k^\prime)\frac{d^3 l^\prime d^3 k^\prime}{(2\pi)^3 E_{l^\prime}E_{k^\prime}}, \label{f:crosssec}
\end{align}
where  $\alpha_{\rm em}$ is the fine structure constant. The $A_W$ factor is calculated as
\begin{align}
  A_W=\frac{Q^4}{\left[(Q^2+M_W^2)^2+\Gamma_W^2M_W^2\right]16\sin^4\theta_W},
\end{align}
where $M_W, \Gamma_W$ indcate the mass and width of $W$ boson, respectively, and the $\theta_W$ denotes the weak mixing angle. 
The leptonic tensor is given by
\begin{align}
  L_{\mu\nu}(l,l')=2\left(l_\mu l'_\nu+l_\nu l'_\mu -g_{\mu\nu} l\cdot l'\right)+2i\lambda_n \varepsilon_{\mu\nu l l'}, \label{f:leptonicten}
\end{align}
where $\lambda_n$ is introduced for convenience. For neutrino production and anti-neutrino production,  $\lambda_n$ is $-1$ and $1$, respectively. The hadronic tensor is given by
\begin{align}
  W^{\mu\nu}(q,p,k^\prime)& = \sum_X (2\pi)^3\delta^4(p + q - k^\prime- p_X) \nonumber \\
  &\times\langle N,S |J^\mu(0)|k^\prime;X\rangle \langle k^\prime;X |J^\nu(0)| N,S \rangle.  \label{f:wgg}
\end{align}
The weak current is defined as $J^\mu(0)=\bar{\psi}(0)\Gamma^\mu \psi(0)$, where $\Gamma^\mu=\gamma^\mu(c_V^q-c_A^q\gamma^5)$ is the interaction vertex, with the weak couplings $c_V^q$ and $c_A^q$. 
For charged-current weak interaction, both couplings take the value $c_V^q=c_A^q=1$.
We keep these symbols in the following context for convenience.
It is also convenient to consider the  $k_T^\prime$-dependent cross section, i.e.,
\begin{align}
  d\sigma = \frac{\alpha_{\rm{em}}^2}{sQ^4}A_{W} L_{\mu\nu}(l,\lambda_n, l^\prime)W^{\mu\nu}(q,p,S,k_T^\prime) \frac{d^3 l^\prime d^2 k_T^\prime}{E_{l^\prime}}, \label{f:crosssection}
\end{align}
where the $k^{\prime}_z$-integrated  hadronic tensor is given by
\begin{align}
W^{\mu\nu}(q,p,S,k_T^\prime) = \int \frac{dk_z^\prime}{(2\pi)^32E_{k^\prime}} W^{\mu\nu}(q,p,S,k^\prime). \label{f:hadronzz}
\end{align}

Following the convention of investigating the neutrino-jet correlations in the $eN$ collinear frame, we define the sum of the momenta of the scattered neutrino (antineutrino) and the jet as $j=l^\prime+k^\prime$. 
In this collinear frame, we have 
\begin{align}
 \vec j_T = \vec l_T^\prime + \vec k_T^\prime =  \vec l_T^\prime + \vec k_T + \vec q_T =\vec k_T, \label{f:jetk}
\end{align}
if higher order gluon radiations are neglected. In other words, the transverse momentum $ j_T$ equals to the intrinsic transverse momentum $k_T$ of a quark in the nucleon (nucleus). 
Therefore, the cross section can be rewritten as
\begin{align}
\frac{d\sigma}{d\eta d^2l'_T d^2 j_T} = \frac{\alpha_{\rm em}^2}{s Q^4} A_W
L_{\mu\nu}(l,\lambda_n, l^\prime) W^{\mu\nu}(q,p,S,j_T), \label{f:crossjperp}
\end{align}
where $\eta$ is the rapidity of the scattered lepton. Here we have used the relation $d\eta= dl'_z /E_{l'}$.

\begin{figure}
  \centering
  \includegraphics[width=0.8\linewidth]{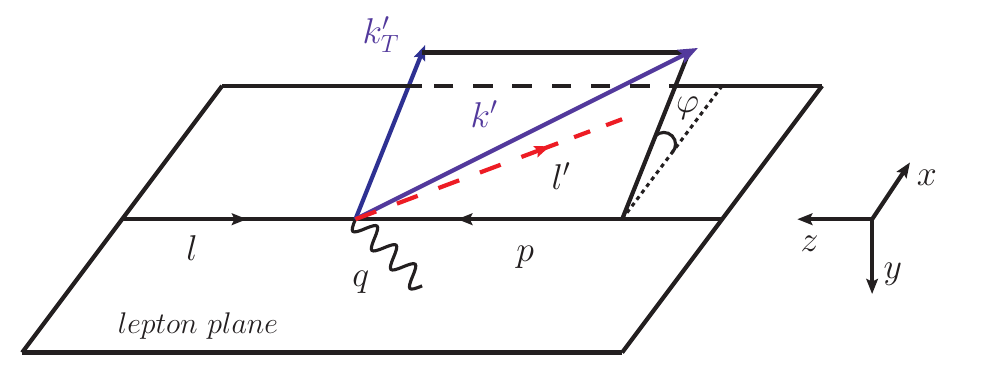}
  \caption{Illustration of the SIDIS process of the jet productions in the $eN$ collinear frame.}\label{fig:gammal}
\end{figure}

As illustrated in Fig.~\ref{fig:gammal}, the $eN$ collinear frame is defined such that the target moves along the $+z$ direction, while the incoming lepton travels along the $-z$ direction. The scattered (anti-)neutrino lies in the $xoz$ plane which is known as the lepton plane.  
In this frame, we use the light-cone unit vectors $\bar t^\mu=(1, 0, \vec 0_T)$ and $t^\mu=(0,  1, \vec 0_T)$ to decompose the relevant momenta, and they satisfy $\bar t^2=t^2=0$ and $\bar t\cdot t=1$. Therefore, we have
\begin{align}
  & p^\mu=p^+ \bar t^\mu + p^- t^\mu, \label{f:pmudef}\\
  & l^\mu =l^+ \bar t^\mu + l^- t^\mu, \label{f:lmudef}
\end{align}
where the plus (minus) components are defined as $p^\pm=\frac{1}{\sqrt{2}}(p^0\pm p^3)$, and similarly for $l^\pm$.
Up to $\mathcal{O}(1/Q^2)$, the momenta can be approximated as $p^\mu \approx p^+ \bar t^\mu$ and $l^\mu \approx l^- t^\mu$.
Correspondingly, the light-cone  vector in this frame can be defined as $\bar t^\mu=p^\mu/ p^+$ and $ t^\mu = l^\mu /l^-$.
Therefore, the relevant momenta can be parametrized as
\begin{align}
  & p^\mu=\left(p^+, 0, \vec 0_T \right), \label{f:pmu}\\
  & l^\mu =\left(0, \frac{Q^2}{2xyp^+}, \vec 0_T \right), \label{f:lmu} \\
  & l^{\prime\mu}=\left(xyp^+, \frac{(1-y)Q^2}{2xyp^+}, Q\sqrt{1-y}, 0\right), \label{f:lpmu}\\ 
  & q^\mu=\left(-xyp^+, \frac{Q^2}{2xp^+}, -Q\sqrt{1-y}, 0\right). \label{f:qmu}
\end{align}
Since the transverse momentum $ j_T$ equals to the intrinsic transverse momentum $k_T$ of a quark in the nucleon in the  $eN$ collinear frame, we parameterize them as 
\begin{align}
 j_T^\mu= k_T^\mu = |k_T|(0,0, \cos\varphi, \sin\varphi), \label{f:kmu}
\end{align}
and we do not distinguish them in the following context.
The polarization vector of the nucleon is described by the helicity $\lambda_N$ and the transverse polarization vector $S_T^\mu$, where $S_T^\mu$ can be decomposed as
\begin{align}
& S_T^\mu = |S_T| \left( 0,0, \cos\varphi_S, \sin\varphi_S \right). \label{f:Smu}
\end{align}
Here we note that the transverse~component is defined with respect to the $z$-direction determined by momenta of the incoming lepton and the target nucleon.

\begin{figure}
  \centering
  \includegraphics[width=0.8\linewidth]{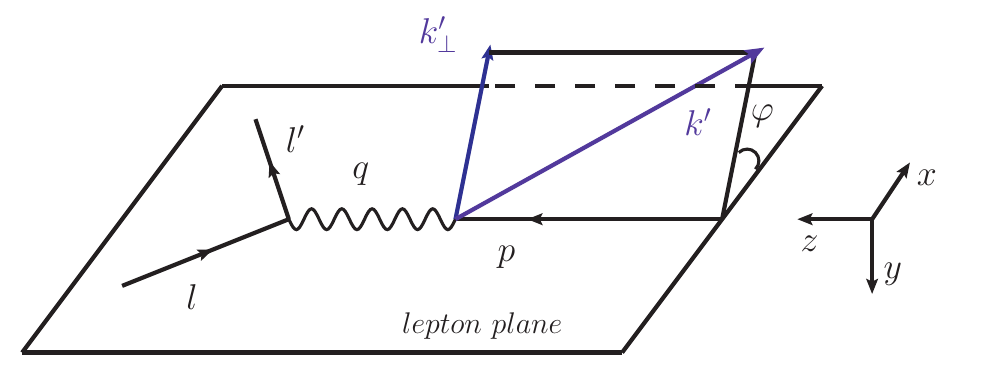}
  \caption{Illustration of the SIDIS process of the jet productions in the $VN$ collinear frame.}\label{fig:gammap}
\end{figure}

Another coordinate frame commonly used in SIDIS is the boson-nucleon (VN) collinear frame, illustrated in Fig.~\ref{fig:gammap}, where the nucleon travels along the $+z$ direction while the intermediate exchange boson moves along the $-z$ direction. 
In this frame, we decompose the relevant momenta using the unit vectors $(\bar{n}, n)$, which are distinguished from $(\bar{t}, t)$ defined in the $eN$ collinear frame. We also have  
\begin{align}
  & p^\mu=\tilde p^+ \bar n^\mu + \tilde p^- n^\mu, \label{f:ptildemudef}\\
  & q^\mu =q^+ \bar n^\mu + q^- n^\mu, \label{f:qmudef}
\end{align}
where the plus component of $p^\mu$ is written as $\tilde p$ to distinguish that in the $eN$ collinear frame. 
Up to $\mathcal{O}(1/Q^2)$, only the large plus component of $p^\mu$ survives.
The light-cone vectors can be defined as $\bar n^\mu=p^\mu/\tilde p^+$ and $n^\mu=(q^\mu+xp^\mu)/q^-$.
Under these conditions, we can parametrize momenta of the involved particles as
\begin{align}
& p^\mu = \left(\tilde p^+,0,\vec 0_\perp \right), \nonumber\\
& q^\mu = \left( -x\tilde p^+, \frac{Q^2}{2x\tilde p^+}, \vec 0_\perp \right),\nonumber\\
& l^\mu = \left( \frac{1-y}{y}x\tilde p^+, \frac{Q^2}{2xy\tilde p^+}, \frac{Q\sqrt{1-y}}{y},0 \right),\nonumber\\
& l^{\prime \mu}=\left( \frac{1}{y}x\tilde p^+, \frac{(1-y)Q^2}{2xy\tilde p^+}, \frac{Q\sqrt{1-y}}{y},0 \right).
\end{align}
The relationships between $(\bar t, t)$ and $(\bar n, n)$ can be expressed as
 \begin{align}
  & \bar t^\mu = \frac{(2-y)}{y}\frac{\tilde p^+}{p^+}\bar n^\mu + \frac{y l_T^\mu + q_T^\mu}{xyp^+},  \label{f:tbart} \\
  & t^\mu = \frac{2(1-y)x^2\tilde p^+ p^+}{y Q^2}\bar n^\mu +\frac{p^+}{\tilde p^+} n^\mu + \frac{2xyp^+}{Q^2} l_T^\mu. \label{f:tt}
 \end{align}
 According to our conventions, the transverse components of $l^\mu$ and $q^\mu$ can be written as 
 \begin{align}
  &l_T=l_x=\frac{Q\sqrt{1-y}}{y}, \\
  & q_T=q_x=-Q\sqrt{1-y},
  \end{align}
 and they lie in the lepton plane or the $x-z$ plane. 
 Under this circumstance, the second term in Eq. (\ref{f:tbart}) vanishes and
 \begin{align}
   p^+ \bar t^\mu =\frac{ (2-y)}{y}\tilde p^+ \bar n^\mu.
 \end{align}
That is to say, $\tilde p^+$ and $p^+$ do not have to be equal.


\section{The cross section in terms of structure functions}\label{sec:structure}

In this section, we perform a general kinematic analysis for this charged-current SIDIS process. We first decompose the hadronic tensor under the limit of kinematic constraints, and then express the differential cross section in terms of structure functions.

Since the hadronic tensor contain the unperturbative information in nucleon, it can not be calculated by perturbative theory.
It is well known that the hadronic tensor must satisfy the conditions required by the hermiticity, gauge invariance, and parity invariance, 
\begin{align}
W^{*\mu\nu}(q, p, S, j_T) &= W^{\nu\mu}(q, p, S, j_T), \\
q_\mu W^{\mu\nu}(q, p, S, j_T) &= q_\nu W^{\mu\nu}(q, p, S, j_T)=0, \\
W^{\mu\nu}(q, p, S, j_T) &= W_{\mu\nu}(q^{\mathcal{P}}, p^{\mathcal{P}}, -S^{\mathcal{P}}, j_T^{\mathcal{P}}),
\end{align}
where the superscript $\mathcal{P}$ indicates a sign flip of all space components, such as $A_\mu^\mathcal{P}=A^\mu$. 
We note here that the parity invariance has no constraint on the hadronic tensor because weak interaction is considered in this process.
With the constraints of hermiticity and gauge invariance, we decompose the hadronic tensor in terms of basic Lorentz tensors (BLTs) multiplied by scalar functions.
One can divide the hadronic tensor into a symmetric and an antisymmetric parts, i.e., $W^{\mu\nu}=W^{S\mu\nu}+iW^{A\mu\nu}$.
The more explicit expression is given by
\begin{align}
W^{S\,\mu\nu} &= \sum_{\sigma, i} W^S_{\sigma i} h^{S\,\mu\nu}_{\sigma i} + \sum_{\sigma, i} \tilde{W}^S_{\sigma i} \tilde{h}^{S\,\mu\nu}_{\sigma i},\label{e.WSmunu} \\
W^{A\,\mu\nu} &= \sum_{\sigma, i} W^A_{\sigma i} h^{A\,\mu\nu}_{\sigma i} + \sum_{\sigma, i} \tilde{W}^A_{\sigma i} \tilde{h}^{A\,\mu\nu}_{\sigma i},\label{e.WAmunu}
\end{align}
where $h^{\mu\nu}_{\sigma i}$'s and $\tilde{h}^{\mu\nu}_{\sigma i}$'s represent the parity conserving and flipping BLTs, respectively.
These coefficients $W_{\sigma i}$'s are scalar functions.
We use the subscript $\sigma$ to specify the spin states of the target particle.

The BLTs can be constructed from the kinematic variables involved in the process.
For unpolarized case, we obtain nine independent BLTs as 
\begin{align}
h^{S\,\mu\nu}_{U_i} &= \left\{ g^{\mu\nu} - \frac{q^\mu q^\nu}{q^2},\; p_q^\mu p_q^\nu,\; j_{Tq}^\mu j_{Tq}^\nu,\; p_q^{\{\mu} j_{Tq}^{\nu\}} \right\}, \label{e.hus}\\
\tilde{h}^{S\,\mu\nu}_{U_i} &= \left\{ \varepsilon^{\{\mu q p j_T} p_q^{\nu\}},\; \varepsilon^{\{\mu q p j_T} j_{Tq}^{\nu\}} \right\}, \\
h_U^{A\,\mu\nu} &= \left\{ p_q^{[\mu} j_{Tq}^{\nu]} \right\}, \\
\tilde{h}^{A\,\mu\nu}_{U_i} &= \left\{ \varepsilon^{\mu\nu q p},\; \varepsilon^{\mu\nu q j_T} \right\},\label{e.hua}
\end{align}
where $p_q^\mu\equiv p^\mu-q^\mu (p\cdot q)/q^2$, which satisfies $p_q\cdot q=0$.
The subscript $U$ indicates the unpolarized part.
We also use the symmetrization and antisymmetrization conventions, i.e., $A^{\{\mu} B^{\mu\}} \equiv A^\mu B^\mu+A^\nu B^\mu$ and $A^{[\mu} B^{\mu]} \equiv A^\mu B^\mu-A^\nu B^\mu$.

As shown in Refs.~\cite{Chen:2016moq,Zhao:2024zpy}, the polarization dependent BLTs can be constructed from multiplying the unpolarized BLTs given in Eqs.~\eqref{e.hus}--\eqref{e.hua} by the polarization dependent scalars or pseudoscalars.
Therefore, we obtain the vector polarized BLTs as follows
\begin{align}
h^{S\,\mu\nu}_{V_i} &= \left\{ \left[(q \cdot S),\; (j_T \cdot S)\right] \tilde{h}^{S\,\mu\nu}_{U_i},\; \varepsilon^{S q p j} h^{S\,\mu\nu}_{U_j} \right\}, \\
\tilde{h}^{S\,\mu\nu}_{V_i} &= \left\{ \left[(q \cdot S),\; (j_T \cdot S)\right] h^{S\,\mu\nu}_{U_i},\; \varepsilon^{S q p j} \tilde{h}^{S\,\mu\nu}_{U_j} \right\}, \\
h^{A\,\mu\nu}_{V_i} &= \left\{ \left[(q \cdot S),\; (j_T \cdot S)\right] \tilde{h}^{A\,\mu\nu}_{U_i},\; \varepsilon^{S q p j} h^{A\,\mu\nu}_U \right\},\\
\tilde{h}^{A\,\mu\nu}_{V_i} &=\left\{ \left[(q \cdot S),\; (j_T \cdot S)\right] h^{A\,\mu\nu}_{U},\; \varepsilon^{S q p j} \tilde{h}^{A\,\mu\nu}_{U j} \right\},\label{e.hVAt}
\end{align}
where the subscript $V$ represent the vector polarized dependence.

Substituting the BLTs shown in Eqs.~\eqref{e.hus}--\eqref{e.hVAt} into Eqs.~\eqref{e.WSmunu} and~\eqref{e.WAmunu}, one can obtain the complete decomposition of hadronic tensor. 
After contracting it with the leptonic tensor, we can express the differential cross section in a general form. Neglecting the complicated calculations, we here only show the final result.
According to the polarization states of the nucleon, the differential cross section can be divided into three parts,
\begin{align}
    \frac{d\sigma}{d\eta d^2l'_T d^2 j_T} =& \frac{\alpha_{\rm em}^2}{Q^4}xy A_W\bigg[
    \mathcal{F}_{U}+\lambda_N \mathcal{F}_{L}+ |S_T|\mathcal{F}_{T}
    \bigg],\label{e.SFs}
\end{align}
where the subscript indicate the polarization states of the target particle.
The explicit form of each part can be expressed in terms of the corresponding structure functions, which are given by
\begin{align}
    \mathcal{F}_{U}=&F_{U}+\sin\varphi F_{U}^{\sin\varphi}+\sin2\varphi F_{U}^{\sin2\varphi}+\cos\varphi F_{U}^{\cos\varphi},\\
    \mathcal{F}_{L}=&F_{L}
    +\sin\varphi F_{L}^{\sin\varphi}
    +\sin2\varphi F_{L}^{\sin2\varphi}+\cos\varphi F_{L}^{\cos\varphi}\nonumber\\
    &+\cos2\varphi F_{L}^{\cos2\varphi}
    +\cos3\varphi F_{L}^{\cos3\varphi},\\
    \mathcal{F}_{T}=&\sin\varphi_S F_{T}^{\sin\varphi_S}
    +\sin(\varphi+\varphi_S)F_{T}^{\sin(\varphi+\varphi_S)}\nonumber\\
    &+\sin(2\varphi+\varphi_S)F_{T}^{\sin(2\varphi+\varphi_S)}
    +\sin(\varphi-\varphi_S)F_{T}^{\sin(\varphi-\varphi_S)}\nonumber\\
    &+\sin(2\varphi-\varphi_S)F_{T}^{\sin(2\varphi-\varphi_S)}
    +\sin(3\varphi-\varphi_S)F_{T}^{\sin(3\varphi-\varphi_S)}\nonumber\\
    &+\cos\varphi_S F_{T}^{\cos\varphi_S}
    +\cos(\varphi+\varphi_S)F_{T}^{\cos(\varphi+\varphi_S)}\nonumber\\
    &+\cos(\varphi-\varphi_S)F_{T}^{\cos(\varphi-\varphi_S)}
    +\cos(2\varphi-\varphi_S)F_{T}^{\cos(2\varphi-\varphi_S)}\nonumber\\
    &+\cos(3\varphi-\varphi_S)F_{T}^{\cos(3\varphi-\varphi_S)}.
 \end{align}
where each term corresponds to a combination of nucleon polarization state and azimuthal modulation.

\section{The calculation in the parton model}\label{sec:partonm}

\subsection{The leading-twist hadronic tensor} 

In the parton model, we calculate the hadronic tensor up to twist-3 level.
In order to show a clear calculation, it is convenient to divide the hadronic tensor into a leading-twist part and a twist-3 part. 
We first begin with the leading-twist contribution. 


At the leading-twist, the hadronic tensor can be obtained by using Eq.~\eqref{f:wgg}. 
According to the factorization, it can be written as
\begin{align}
  W^{\mu\nu}=\frac{ 2E_{k'}}{2p\cdot q}{\rm{Tr}} \left[\hat \Phi^{(0)}(x, k_T) \hat H^{\mu\nu}(q,k)  \right] (2\pi)^3 \delta(q_z+k_z-k_z^\prime).
\end{align}
Integrating over $k'_z$, see Eq.~\eqref{f:hadronzz}, we obtain the $k'_T (j_T)$-dependent hadronic tensor, 
\begin{align}
   W^{\mu\nu}=\frac{1}{2p\cdot q} {\rm{Tr}} \left[\hat \Phi^{(0)}(x, k_T) \hat H^{\mu\nu}(q,k) \right], \label{f:hadronlead}
\end{align}
where the quark-quark correlator is defined as
\begin{align}
   \hat \Phi^{(0)}(x,k_T)&= \int \frac{p^+ d\xi^- d^2 \vec \xi_T}{(2\pi)^3} e^{ixp^+ \xi^- - i\vec k_T \vec\xi_T}\nonumber \\
  &\times\langle N,S| \bar \psi(0) {\cal L}(0,y) \psi(\xi)|N,S \rangle. \label{f:correlator}
\end{align}
The gauge link has been inserted into the quark-quark correlator to keep the gauge invariance. We decompose the correlators as 
\begin{align}
& \hat \Phi^{(0)}= \frac{1}{2}\left[\gamma^\alpha \Phi^{(0)}_\alpha + \gamma^\alpha\gamma_5 \tilde\Phi^{(0)}_\alpha \right], \label{f:hatphi0}
\end{align}
In the jet-production SIDIS process, we only need to consider the $\gamma^\alpha$- and the $\gamma^\alpha\gamma^5$-terms (chiral-even terms) in the parametrization of these correlators, because there is no helicity flips. 
The TMDs are obtained by decomposing the correlator which are given by
\begin{align}
  \Phi^{(0)}_\alpha &=p^+ \bar t_\alpha\Bigl(f_1-\frac{k_T \cdot \tilde S_T}{M}f^\perp_{1T}  \Bigr) +k_{T\alpha} f^\perp  \nonumber\\
  &- M\tilde S_{T\alpha}f_T - \lambda_N \tilde k_{T\alpha} f_L^\perp  -\frac{k_{T\langle\alpha}k_{T\beta\rangle}}{M}  \tilde S_T^\beta f_T^\perp  ,
\label{eq:Xi0Peven}\\
  \tilde\Phi^{(0)}_\alpha &=p^+\bar t_\alpha\Bigl(-\lambda_N g_{1L}+\frac{k_T\cdot S_T}{M}g^\perp_{1T}\Bigr)-\tilde k_{T\alpha}  g^\perp\nonumber\\
  &- M S_{T\alpha}g_T -\lambda_N k_{T\alpha} g_L^\perp + \frac{k_{T\langle\alpha}k_{T\beta\rangle}}{M}  S_T^\beta g_T^\perp .
\label{eq:Xi0Podd}
\end{align}
We have introduced the notation $\tilde{A}^\alpha = \varepsilon_T^{\alpha A} = \varepsilon_T^{\alpha \beta} A_{T\beta}$, where $A$ can be either $k_T$ or $S_T$.
The symmetric traceless tensor is defined as $k_{T\langle\alpha}k_{T\beta\rangle} = k_{T\alpha}k_{T\beta} - \frac{1}{2}g_{T\alpha\beta}k_T^2$. 


The hard part in Eq. (\ref{f:hadronlead}) is abbreviated as
\begin{align}
  \hat H^{\mu\nu}(q,k) =\Gamma^{\mu}(\slashed q+\slashed k) \Gamma^{\nu}. \label{f:hardamp}
\end{align}
In the $eN$ collinear frame, we have the following relationships
\begin{align}
 &k^- \ll k_T \ll q_T \sim q^-,\\
 &q^+ + k^+=(1-y)xp^+.
\end{align}
Neglecting the small components of $k$, we obtain
\begin{align}
  (\slashed q+\slashed k) =(1-y)xp^+ \slashed{\bar t} + q^- \slashed t +\slashed q_T. \label{f:harden}
\end{align}
We notice that the transverse and the plus components also contribute in the $eN$ collinear frame in addition to the minus component. This is important to the requirement of the current conservation of the hadronic tensor. 


When the dust settles, the hadronic tensor at leading-twist can be expressed as
\begin{align}
 W_{t2}^{\mu\nu} =& -\left( c_1^q \tilde g_{T}^{\mu\nu} + ic_3^q \tilde \varepsilon_{T}^{\mu\nu} \right) \Bigl(f_1-\frac{k_T \cdot \tilde S_T}{M}f^\perp_{1T} \Bigr) \nonumber\\
& - \left(c_3^q \tilde g_{T}^{\mu\nu} + i c_1^q \tilde\varepsilon_{T}^{\mu\nu} \right) \Bigl(-\lambda_N g_{1L}+\frac{k_T\cdot S_T}{M}g^\perp_{1T} \Bigr), \label{f:wt2munu}
\end{align}
where the subscript $t2$ denotes leading-twist. The $\tilde g_{T}^{\mu\nu}$ and $\tilde \varepsilon_{T}^{\mu\nu}$ are dimensionless tensors,
\begin{align}
  & \tilde g_{T}^{\mu\nu} =g_{T}^{\mu\nu} -\frac{\vec q_T^2}{(q^-)^2}\bar t^\mu \bar t^\nu -\frac{1}{q^-}q_T^{\{\mu} \bar t^{\nu\}}, \label{f:gperp}\\
  & \tilde \varepsilon_{T}^{\mu\nu} = \varepsilon_{T}^{\mu\nu} +\frac{1}{q^-}\varepsilon^{\mu\nu \bar t q_T}. \label{f:eperp}
\end{align}
It is easily to check that $q_\mu \tilde{g}_T^{\mu\nu}=q_\nu \tilde{g}_T^{\mu\nu}=0$ and $q_\mu \tilde{\varepsilon}_T^{\mu\nu}=q_\nu \tilde{\varepsilon}_T^{\mu\nu}=0$ by using the definitions
\begin{align}
  & g_T^{\mu\nu} = g^{\mu\nu} - \bar{t}^\mu t^\nu - \bar{t}^\nu t^\mu, \\
  & \varepsilon_T^{\mu\nu}= \varepsilon^{\alpha\beta\mu\nu}\bar{t}_\alpha t_\beta.
\end{align}
These equations imply that the hadronic tensor satisfies the current conservation.

\subsection{The twist-3 hadronic tensor}

The twist-3 contributions to the hadronic tensor arise from two sources. One is the quark-quark correlator given in Eq. (\ref{f:correlator}), the other is the quark-gluon-quark correlator whose operator definition is given by
\begin{align}
  \hat{\Phi}_{\rho}^{(1)}\left(x, k_{T}\right) &= \int \frac{p^{+} d y^{-} d^{2} y_{T}}{(2 \pi)^{3}} e^{i x p^+ y^- - i \vec{k}_T \cdot \vec{y}_T} \nonumber\\
   &\times \langle p,S| \bar{\psi}(0) D_{T \rho}(0) {\cal L}(0, y) \psi(y)| p,S\rangle, \label{f:Phi1}
\end{align}
where $D_\rho(y) = -i\partial_\rho + g A_\rho(y)$ is the covariant derivative. We decompose $\hat{\Phi}_{\rho}^{(1)}$ as
\begin{align}
& \hat \Phi_\rho^{(1)}= \frac{1}{2}\left[\gamma^\alpha \Phi_{\rho\alpha}^{(1)} + \gamma^\alpha\gamma_5 \tilde\Phi_{\rho\alpha}^{(1)} \right].
\end{align}
Similar to Eqs. (\ref{eq:Xi0Peven}) and (\ref{eq:Xi0Podd}), the coefficient functions are decomposed as 
\begin{align}
  \Phi^{(1)}_{\rho\alpha}&=p^+\bar t_\alpha\Bigl[k_{T\rho} f^\perp_d- M\tilde S_{T\rho}f_{dT} \nonumber\\
  & -\lambda_N \tilde k_{T\rho} f_{dL}^\perp -\frac{k_{T\langle\rho}k_{T\beta\rangle}}{M} \tilde S_T^\beta f_{dT}^\perp \Bigr], \label{eq:Xi1Peven} \\
  \tilde \Phi^{(1)}_{\rho\alpha}&=ip^+\bar t_\alpha\Bigl[\tilde k_{T\rho}g^\perp_d+ MS_{T\rho}g_{dT}\nonumber\\
  & +\lambda_N k_{T\rho} g_{dL}^\perp  - \frac{k_{T\langle\rho}k_{T\beta\rangle}}{M} S_T^\beta g_{dT}^\perp  \Bigr],
\label{eq:Xi1Podd}
\end{align}
where the subscript $d$ is used to denote twist-3 TMDs defined via the quark-gluon-quark correlator. 

Since the complete calculation of the twist-3 hadronic tensor is complicated, we only present the main steps by taking the $f^\perp$ and $g^\perp$ terms for example. 

First of all, we calculate the contributions from the quark-quark correlator. Inserting the twist-3 TMDs \eqref{eq:Xi0Peven}--\eqref{eq:Xi0Podd} and the hard part \eqref{f:hardamp} into Eq. (\ref{f:hadronlead}), we obtain
\begin{align}
  W^{\mu\nu}_{t3,q} &=\bigg[ c_1^q \left(k_T^{\{\mu} t^{\nu\}}q^- + k_T^{\{\mu} \bar t^{\nu\}}(1-y)xp^+ + k_T^{\{\mu} q_T^{\nu\}}- g^{\mu\nu}k_T\cdot q_T\right)   \nonumber\\
  & +ic_3^q\left(\tilde k_T^{[\mu} t^{\nu]}q^- -\tilde k_T^{[\mu} \bar t^{\nu]}(1-y)xp^+ - \bar t^{[\mu} t^{\nu]}\varepsilon_T^{ k q}\right) \bigg] \frac{ f^\perp}{p\cdot q}\nonumber\\
 &+\bigg[- c_3^q \left(\tilde k_T^{\{\mu} t^{\nu\}}q^- + \tilde k_T^{\{\mu} \bar t^{\nu\}}(1-y)xp^+ + \tilde k_T^{\{\mu} q_T^{\nu\}}- g^{\mu\nu}\varepsilon_T^{ k q}\right)  \nonumber\\
  & +ic_1^q\left( k_T^{[\mu} t^{\nu]}q^- -k_T^{[\mu} \bar t^{\nu]}(1-y)xp^+ - \bar t^{[\mu} t^{\nu]}k_T\cdot q_T\right) \bigg]\frac{ g^\perp }{p\cdot q}.
  \label{f:wt31}
\end{align}
where the subscript $q$ denotes hadronic tensor from the quark-quark correlator. Because of the incompleteness of the twist-3 hadronic tensor, this expression does not satisfy the current conservation, i.e., $q_\mu W^{\mu\nu}_{t3,q} \neq 0$.

Next we calculate the contributions from the quark-gluon-quark correlator.
There are two parts, $W^{\mu\nu}_{t3, L}$ and $W^{\mu\nu}_{t3, R}$, which satisfy $ W^{\mu\nu}_{t3, L}= (W^{\nu\mu}_{t3, R})^*$. The indices $L$ and $R$ denote the left-cut and the right-cut, respectively \cite{Liang:2006wp}. According to the operator definition of the  hadronic tensor, we have
\begin{align}
  W^{\mu\nu}_{t3, L}=\frac{1}{2p\cdot q} {\rm{Tr}} \left[\hat \Phi_\rho^{(1)}(x, k_T) \hat H^{\mu\nu,\rho}(q, k_1, k_2) \right], \label{f:hadronnext}
\end{align}
where $\hat H^{\mu\nu,\rho}$ is the hard scattering amplitude,
\begin{align}
  \hat H^{\mu\nu, \rho} =\Gamma^{\mu,q}\frac{\slashed k_2 +\slashed q}{(k_2+q)^2}\gamma^\rho(\slashed k_1+\slashed q) \Gamma^{\nu,q}. \label{f:hardamp1}
\end{align}
Substituting Eqs. (\ref{eq:Xi1Peven}), (\ref{eq:Xi1Podd}) and  (\ref{f:hardamp1}) into (\ref{f:hadronnext}) gives 
\begin{align}
  W^{\mu\nu}_{t3, L}& =  \left[c_1^q \left(\bar t^\mu \bar t^\nu \frac{k_T\cdot q_T}{q^-} - k_T^\nu \bar t^\mu\right) \right. \nonumber\\
  &\left. \ \ - ic_3^q  \left(\bar t^\mu \bar t^\nu \frac{\varepsilon_T^{qk}}{q^-} - \tilde k_T^\nu \bar t^\mu\right)  \right]\frac{p^+}{p\cdot q} f^\perp_d \nonumber \\
  & -  \left[c_1^q \left(\bar t^\mu \bar t^\nu \frac{k_T\cdot q_T}{q^-} - k_T^\nu \bar t^\mu\right) \right. \nonumber \\
  &\left.  \ \ - ic_3^q  \left(\bar t^\mu \bar t^\nu \frac{\varepsilon_T^{qk}}{q^-} - \tilde k_T^\nu \bar t^\mu\right)  \right]\frac{p^+}{p\cdot q} g^\perp_d.   \label{f:wt3l}
\end{align}
We note that twist-3 TMDs marked with subscript $d$ can be related to those without the subscript through the following relation \cite{Chen:2020ugq}
\begin{align}
f_{d S}^{K}-g_{d S}^{K}=-x\left(f_{S}^{K}-i g_{S}^{K}\right),\label{f:formulaEOM}
\end{align}
where the superscript $K$ can be $\perp$ and the subscript $S$ can be $L$ and $T$. 
Using this relation, Eq. (\ref{f:wt3l}) can be rewritten as 
\begin{align}
  W^{\mu\nu}_{t3, L} =&-\left[c_1^q \left( \bar t^\mu \bar t^\nu \frac{k_T \cdot q_T}{q^-}- k_T^\nu \bar t^\mu \right) \right. \nonumber \\
   & \left. - ic_3^q \left(\tilde{k}^\nu_T \bar t^\mu - \bar t^\mu \bar t^\nu \frac{\varepsilon_T^{qk}}{q^-}\right)\right]\frac{xp^+}{p\cdot q}f^\perp \nonumber \\
   &+ \left[ic_1^q \left( \bar t^\mu \bar t^\nu \frac{k_T \cdot q_T}{q^-}- k_T^\nu \bar t^\mu \right) \right. \nonumber \\
   & \left. +c_3^q \left(\tilde{k}^\nu_T \bar t^\mu - \bar t^\mu \bar t^\nu \frac{\varepsilon_T^{qk}}{q^-}\right)\right]\frac{xp^+}{p\cdot q}g^\perp. \label{f:wmunufgd}
\end{align}
Note that the complete twist-3 hadronic tensor from the quark-gluon-quark correlator is given by the sum of $W^{\mu\nu}_{t3, L}$ and $W^{\mu\nu}_{t3, R}$,
\begin{align}
    & W^{\mu\nu}_{t3, L}+  W^{\mu\nu}_{t3, R} \nonumber\\
    = & - \left[c_1^q \left(\bar t^\mu \bar t^\nu \frac{2x p^+}{q^-} k_T\cdot q_T - xp^+ k_T^{\{\mu} \bar t^{\nu\}}\right) - ic_3^q  x p^+\tilde k_T^{[\mu} \bar t^{\nu]} \right]\frac{ f^\perp}{p\cdot q} \nonumber \\
  & +\left[c_3^q  \left(\bar t^\mu \bar t^\nu \frac{2x p^+}{q^-}\varepsilon_T^{kq} - xp^+ \tilde  k_T^{\{\mu} \bar t^{\nu\}}\right)+ic_1^qx p^+ k_T^{[\mu} \bar t^{\nu]}   \right]\frac{g^\perp}{p\cdot q}.  \label{f:wt3lr}
\end{align}

To obtain the complete twist-3 hadronic tensor, which satisfies the current conservation, one is supposed to sum Eqs. (\ref{f:wt31}) and (\ref{f:wt3lr}) together. Finally, we have
\begin{align}
  W^{\mu\nu}_{t3} &=\frac{ f^\perp}{p\cdot q}h^{\mu\nu}_1 +\frac{g^\perp}{p\cdot q}h^{\mu\nu}_3 +\cdots, 
\end{align}
where the tensors $h^{\mu\nu}_1$ and $h^{\mu\nu}_3$ are given by
\begin{widetext}
\begin{align}
   h_1^{\mu\nu}=& +c_1^q \bigg[ k_T^{\{\mu} t^{\nu\}}q^- + k_T^{\{\mu} \bar t^{\nu\}}(2-y)xp^+ + k_T^{\{\mu} q_T^{\nu\}}- \bigg( g^{\mu\nu}+\bar t^\mu \bar t^\nu \frac{2x p^+}{q^-} \bigg) k_T\cdot q_T \bigg]  \nonumber\\
  & + ic_3^q \bigg[ \tilde k_T^{[\mu} t^{\nu]}q^- -\tilde k_T^{[\mu} \bar t^{\nu]}q^+ - \bar t^{[\mu} t^{\nu]}\varepsilon_T^{qk}\bigg],  \label{f:hmunu1}\\
  h_3^{\mu\nu}=&-c_3^q \bigg[\tilde k_T^{\{\mu} t^{\nu\}}q^- + \tilde k_T^{\{\mu} \bar t^{\nu\}}(2-y)xp^+ + \tilde k_T^{\{\mu} q_T^{\nu\}}- \bigg( g^{\mu\nu}+\bar t^\mu \bar t^\nu \frac{2x p^+}{q^-} \bigg) \varepsilon_T^{q k}\bigg]  \nonumber\\
   & + ic_1^q \bigg[ k_T^{[\mu} t^{\nu]}q^- - k_T^{[\mu} \bar t^{\nu]}q^+ - \bar t^{[\mu} t^{\nu]}k_T\cdot q_T \bigg].  \label{f:hmunu3}
\end{align}
Here we only show the $f^\perp$ and $g^\perp$ terms. 
By performing the steps presented above, the twist-3 hadronic tensor including the previously omitted terms can be written as 
\begin{align}
  W^{\mu\nu}_{t3} &=\frac{ f^\perp}{p\cdot q}h^{\mu\nu}_1 + \frac{ \lambda_N f^\perp_L}{p\cdot q}h^{\mu\nu}_2+\frac{g^\perp}{p\cdot q}h^{\mu\nu}_3 + \frac{\lambda g^\perp_L}{p\cdot q} h^{\mu\nu}_4 \nonumber \\
  &+ \frac{Mf_T}{p\cdot q}h_5^{\mu\nu} + \frac{f^\perp_T}{p\cdot q}h_6^{\mu\nu} + \frac{Mg_T}{p\cdot q}h_7^{\mu\nu} + \frac{g^\perp_T}{p\cdot q}h_8^{\mu\nu},  \label{f:wt3}
\end{align}
where $h^{\mu\nu}_{2, 4-8}$ are defined as
\begin{align}
  h_2^{\mu\nu}=&-c_1^q \bigg[\tilde k_T^{\{\mu} t^{\nu\}}q^- + \tilde k_T^{\{\mu} \bar t^{\nu\}}(2-y)xp^+ + \tilde k_T^{\{\mu} q_T^{\nu\}}- \bigg( g^{\mu\nu}+\bar t^\mu \bar t^\nu \frac{2x p^+}{q^-} \bigg) \varepsilon_T^{q k}\bigg]  \nonumber\\
  & + ic_3^q \bigg[ k_T^{[\mu} t^{\nu]}q^- - k_T^{[\mu} \bar t^{\nu]}q^+ - \bar t^{[\mu} t^{\nu]}k_T\cdot q_T \bigg],  \label{f:hmunu2}\\
    h_4^{\mu\nu}=&-c_3^q \bigg[ k_T^{\{\mu} t^{\nu\}}q^- + k_T^{\{\mu} \bar t^{\nu\}}(2-y)xp^+ + k_T^{\{\mu} q_T^{\nu\}}- \bigg( g^{\mu\nu}+\bar t^\mu \bar t^\nu \frac{2x p^+}{q^-} \bigg) k_T\cdot q_T \bigg]  \nonumber\\
   & - ic_1^q \bigg[ \tilde k_T^{[\mu} t^{\nu]}q^- -\tilde k_T^{[\mu} \bar t^{\nu]}q^+ - \bar t^{[\mu} t^{\nu]}\varepsilon_T^{qk}\bigg],  \label{f:hmunu4}\\
    h_5^{\mu\nu}=&+\Bigg\{-c_1^q \bigg[\tilde S_T^{\{\mu} t^{\nu\}}q^- + \tilde S_T^{\{\mu} \bar t^{\nu\}}(2-y)xp^+ + \tilde S_T^{\{\mu} q_T^{\nu\}}- \bigg( g^{\mu\nu}+\bar t^\mu \bar t^\nu \frac{2x p^+}{q^-} \bigg) \varepsilon_T^{  q S}\bigg]  \nonumber\\
  &\quad + ic_3^q \bigg[ S_T^{[\mu} t^{\nu]}q^- - S_T^{[\mu} \bar t^{\nu]}q^+ - \bar t^{[\mu} t^{\nu]}S_T\cdot q_T \bigg]\Bigg\},  \label{f:hmunu5}\\
   h_6^{\mu\nu}=& -\Bigg\{+c_1^q \bigg[ k_T^{\{\mu} t^{\nu\}}q^- + k_T^{\{\mu} \bar t^{\nu\}}(2-y)xp^+ + k_T^{\{\mu} q_T^{\nu\}}- \bigg( g^{\mu\nu}+\bar t^\mu \bar t^\nu \frac{2x p^+}{q^-} \bigg) k_T\cdot q_T \bigg]  \nonumber\\
  &\quad + ic_3^q \bigg[ \tilde k_T^{[\mu} t^{\nu]}q^- -\tilde k_T^{[\mu} \bar t^{\nu]}q^+ - \bar t^{[\mu} t^{\nu]}\varepsilon_T^{qk}\bigg] \Bigg\}\frac{\varepsilon_T^{kS}}{M} \\
  &-\Bigg\{-c_1^q \bigg[\tilde S_T^{\{\mu} t^{\nu\}}q^- + \tilde S_T^{\{\mu} \bar t^{\nu\}}(2-y)xp^+ + \tilde S_T^{\{\mu} q_T^{\nu\}}- \bigg( g^{\mu\nu}+\bar t^\mu \bar t^\nu \frac{2x p^+}{q^-} \bigg) \varepsilon_T^{  q S}\bigg]  \nonumber\\
  &\quad + ic_3^q \bigg[ S_T^{[\mu} t^{\nu]}q^- - S_T^{[\mu} \bar t^{\nu]}q^+ - \bar t^{[\mu} t^{\nu]}S_T\cdot q_T \bigg]\Bigg\} \frac{k_T^2}{2M},   \label{f:hmunu6}\\
    h_7^{\mu\nu}=&+\Bigg\{-c_3^q \bigg[ S_T^{\{\mu} t^{\nu\}}q^- +  S_T^{\{\mu} \bar t^{\nu\}}(2-y)xp^+ +  S_T^{\{\mu} q_T^{\nu\}}- \bigg( g^{\mu\nu}+\bar t^\mu \bar t^\nu \frac{2x p^+}{q^-} \bigg)S_T\cdot q_T\bigg]  \nonumber\\
  & \quad- ic_1^q \bigg[ \tilde S_T^{[\mu} t^{\nu]}q^- -\tilde S_T^{[\mu} \bar t^{\nu]}q^+ - \bar t^{[\mu} t^{\nu]}\varepsilon_T^{  q S} \bigg]\Bigg\},  \label{f:hmunu7}\\
   h_8^{\mu\nu}=& +\Bigg\{+c_3^q \bigg[ k_T^{\{\mu} t^{\nu\}}q^- + k_T^{\{\mu} \bar t^{\nu\}}(2-y)xp^+ + k_T^{\{\mu} q_T^{\nu\}}- \bigg( g^{\mu\nu}+\bar t^\mu \bar t^\nu \frac{2x p^+}{q^-} \bigg) k_T\cdot q_T \bigg]  \nonumber\\
  &\quad + ic_1^q \bigg[ \tilde k_T^{[\mu} t^{\nu]}q^- -\tilde k_T^{[\mu} \bar t^{\nu]}q^+ - \bar t^{[\mu} t^{\nu]}\varepsilon_T^{qk}\bigg] \Bigg\}\frac{k_T\cdot S_T}{M} \\
  &-\Bigg\{+c_3^q \bigg[ S_T^{\{\mu} t^{\nu\}}q^- +  S_T^{\{\mu} \bar t^{\nu\}}(2-y)xp^+ +  S_T^{\{\mu} q_T^{\nu\}}- \bigg( g^{\mu\nu}+\bar t^\mu \bar t^\nu \frac{2x p^+}{q^-} \bigg)S_T\cdot q_T\bigg]  \nonumber\\
  & \quad+ic_1^q \bigg[ \tilde S_T^{[\mu} t^{\nu]}q^- -\tilde S_T^{[\mu} \bar t^{\nu]}q^+ - \bar t^{[\mu} t^{\nu]}\varepsilon_T^{  q S} \bigg]\Bigg\}\frac{k_T^2}{2M}.   \label{f:hmunu8}
\end{align}
\end{widetext}
We notice that the complete twist-3 hadronic tensor in Eq.~\eqref{f:wt3} satisfies the current conservation, $q_\mu \tilde W^{\mu\nu}_{t3} = q_\nu \tilde W^{\mu\nu}_{t3} = 0$.
Although the $h$-tensors appear somewhat complicated, they share similar structures and will lead to a simple expression for the differential cross section.

\subsection{The cross section in the parton model}\label{sec:crossection}

Contracting the hadronic tensor obtained in Eq. (\ref{f:wt3}) with the leptonic tensor, the differential cross section can be expressed in terms of TMDs in the parton model. 
At the leading-twist approximation, the contractions are given by
\begin{align}
  & L_{\mu\nu} \left(c_1^q \tilde g_T^{\mu\nu} + ic_3^q \tilde\varepsilon_T^{\mu\nu}\right) = -\frac{2Q^2}{y^2}T_0^q(y), \\
  & L_{\mu\nu} \left(c_3^q \tilde g_T^{\mu\nu} + ic_1^q \tilde\varepsilon_T^{\mu\nu}\right) = -\frac{2Q^2}{y^2}T_1^q(y),
\end{align}
where $T$-functions are defined as
\begin{align}
  & T_0^q(y) =  c_1^q A(y) - \lambda_n c_3^q C(y), \nonumber\\
  & T_1^q(y) =  c_3^q A(y) - \lambda_n c_1^q C(y). \label{eq:T0T1}
\end{align}
Here $A(y)=y^2-2y+2$ and $C(y)=y(2-y)$.
Since $c_1^q = c_3^q = 2$ in the charged-current interaction, it follows that $T_0^q(y) = T_1^q(y)$. For simplicity, we define $T^q(y) \equiv T_0^q(y) = T_1^q(y)$ in the following.
Therefore, the leading-twist cross section for the jet-production SIDIS of electrion in the $eN$ collinear frame can be expressed as
\begin{align}
  d\tilde{\sigma}_{t2}
  &= \frac{\alpha_{\rm em}^2 }{y Q^4} A_W T^q(y) 2x\Biggl\{  f_1 - \lambda_N g_{1L} \nonumber\\
  &+|S_T|k_{T M}\Big[\sin(\varphi-\varphi_S) f^\perp_{1T} -\cos(\varphi-\varphi_S) g^\perp_{1T}\Big] \Biggr\}, \label{f:crosslead}
\end{align}
where $d\tilde{\sigma}_{t2}=d\sigma_{t2}/d\eta d^2l'_T d^2 j_T$ and
$k_{T M}=|k_T|/M$. The subscript $t2$ denotes the leading-twist contribution.

Similarly, we can calculate the differential cross section at twist-3. 
For conciseness, we only show contractions of the leptonic tensor with $h^{\mu\nu}_{1}$ and $h^{\mu\nu}_{3}$,
\begin{align}
  & L_{\mu\nu}\cdot h_1^{\mu\nu} = -\frac{2Q^3}{y^2}|k_T| T_2^q(y)\cos\varphi, \\
  & L_{\mu\nu}\cdot h_3^{\mu\nu} = -\frac{2Q^3}{y^2}|k_T| T_3^q(y)\sin\varphi.
\end{align}
Other contractions have the same forms. Here the $T$-functions are defined as
\begin{align}
  & T_2^q(y) = c_1^q B(y) -\lambda_n c_3^q D(y), \nonumber\\
  & T_3^q(y) = c_3^q B(y) -\lambda_n c_1^q D(y), \label{eq:T2T3}
\end{align}
with $B(y)=(2-y^2)\sqrt{1-y}$ and $D(y)=y^2\sqrt{1-y}$.
After simple calculations, we write down the differential cross section at twist-3,
\begin{align}
  d\tilde{\sigma}_{t3}&= -\frac{\alpha_{\rm em}^2 }{y Q^4}A_W 4x^2 \kappa_M \tilde{T}^q(y) \Biggl\{k_{T M}  \left[\cos\varphi f^\perp +\sin\varphi g^\perp \right]\nonumber\\
  & \quad +\lambda_N  k_{T M}\left[\sin\varphi f^\perp_{L}-\cos\varphi g^\perp_{L} \right] \nonumber \\
  &+ |S_T|\Big[\sin\varphi_S f_T +\sin(2\varphi-\varphi_S)\frac{k_{T M}^2}{2}f_T^\perp \nonumber \\
  &\quad +\cos\varphi_S  g_T -\cos(2\varphi-\varphi_S) \frac{k_{T M}^2}{2}g_T^\perp \Big]\Biggr\},\label{e.crosstwist3}
\end{align}
 where $d\tilde{\sigma}_{t3}=d\sigma_{t3}/d\eta d^2l'_T d^2 j_T$.
 $\kappa_M=M/Q$ is the twist suppression factor. $\tilde{T}^q(y)$ is defined as $\tilde{T}^q(y)\equiv T_2^q(y)=T_3^q(y)$ because of $c_1^q=c_3^q=2$.
We note that in the electron scattering process, $U$-type quarks ($u, \bar{d}, \bar{s}, \cdots$) contribute to the cross section while in the positron scattering process, $D$-type quarks ($d, s, \bar{u}, \cdots$) contribute to the cross section.

\section{Measurable quantities} \label{sec:measurable}
\subsection{The structure functions}

Comparing the cross section in terms of structure functions in Eq.~\eqref{e.SFs} and the parton model results in Eqs.~\eqref{f:crosslead} and \eqref{e.crosstwist3}, one can obtain the results of the structure functions in the parton model. In the following, we take the SIDIS of electron as an example to present the results. 
There are four nonzero structure functions at the leading twist,
\begin{align}
    &F_{U}=2\frac{T^q(y)}{y^2} f_1, \label{f:FU}\\
    &F_{L}=-2 \frac{T^q(y)}{y^2} g_{1L},\\
    &F_{T}^{\sin(\varphi-\varphi_S)}=2 \frac{T^q(y)}{y^2} k_{TM} f_{1T}^\perp,\\
    &F_{T}^{\cos(\varphi-\varphi_S)}=-2 \frac{T^q(y)}{y^2} k_{TM} g_{1T}^\perp.
\end{align}
At twist-3, we obtain eight nonzero structure functions, 
\begin{align}
    &F_{U}^{\cos\varphi}=-4 \frac{\tilde{T}^q(y)}{y^2}x\kappa_M k_{TM} f^\perp,\\
    &F_{U}^{\sin\varphi}=-4\frac{\tilde{T}^q(y)}{y^2}x\kappa_M k_{TM} g^\perp,\\
    &F_{L}^{\sin\varphi}=-4\frac{\tilde{T}^q(y)}{y^2}x\kappa_M k_{TM} f_L^\perp,\\
    &F_{L}^{\cos\varphi}=4 \frac{\tilde{T}^q(y)}{y^2}x\kappa_M k_{TM} g_L^\perp,\\
    &F_{T}^{\sin\varphi_S}=-4 \frac{\tilde{T}^q(y)}{y^2}x\kappa_M f_T,\\
    &F_{T}^{\sin(2\varphi-\varphi_S)}=-4\frac{\tilde{T}^q(y)}{y^2}x\kappa_M \frac{k_{TM}^2}{2} f_T^\perp,\\
    &F_{T}^{\cos\varphi_S}=-4\frac{\tilde{T}^q(y)}{y^2}x\kappa_M g_T,\\
    &F_{T}^{\cos(2\varphi-\varphi_S)}=4 \frac{\tilde{T}^q(y)}{y^2}x\kappa_M \frac{k_{TM}^2}{2} g_T^\perp. \label{f:FTcos2}
\end{align}
According to the results above, we find that results of the structure functions including only leading-twist functions exhibit the dependence on the even number of $\varphi$ and $\varphi_{S}$, while the structure functions at twist-3 correspond to the dependence of the odd number of $\varphi$ and $\varphi_S$.
We can in principle utilize these nonvanishing structure functions to study the coupling of the weak interactions.
 


\subsection{The azimuthal asymmetries}
For the jet-production SIDIS, FFs are not involved. Therefore, (TMD) PDFs are the only unknown quantities related to azimuthal asymmetries. From the cross sections shown before we notice that azimuthal asymmetries refer to azimuthal asymmetries of $k_T$, or azimuthal asymmetries of $j_T$, since $\vec{j}_T=\vec{k}_T$ in the $eN$ collinear from.  One therefore can measure $j_T$ to determine $k_T$. 
We here use the Trento convention to define the azimuthal asymmetry. For example, we define
\begin{align}
  \langle \sin\varphi \rangle_{U}=\frac{\int d\tilde\sigma \sin\varphi d\varphi}{\int d\tilde\sigma d\varphi},
\end{align}
for the unpolarized or longitudinally polarized target case, and
\begin{align}
  \langle \sin(\varphi-\varphi_S) \rangle_{T}=\frac{\int d\tilde\sigma \sin(\varphi-\varphi_S)d\varphi d\varphi_S}{\int d\tilde\sigma d\varphi d\varphi_S},
\end{align}
for the transversely polarized target case.

At the leading-twist, there are two polarization dependent azimuthal asymmetries which are given by
\begin{align}
 & \langle \sin(\varphi-\varphi_S) \rangle_{T} = \frac{k_{T M}}{2} \frac{ f^\perp_{1T}}{ f_1}, \\ 
 & \langle \cos(\varphi-\varphi_S) \rangle_{T} = -\frac{k_{T M} }{2} \frac{g^\perp_{1T}}{f_1}.
\end{align}
$\langle \sin(\varphi-\varphi_S) \rangle_{T}$ is the famous Sivers asymmetry or single transverse spin asymmetry. In the jet-production charged-current SIDIS, it is just a ratio of Sivers function $f_{1T}^\perp$ \cite{Sivers:1989cc,Sivers:1990fh} and $f_1$. In addition, we also have eight twist-3 azimuthal asymmetries. They are given by
\begin{align}
  & \langle \cos\varphi \rangle_{U} = -x\kappa_M k_{T M} \frac{\tilde T^q(y)}{T^q(y)}\frac{f^\perp}{f_1}, \\
  & \langle \sin\varphi \rangle_{U} = -x\kappa_M k_{T M} \frac{\tilde T^q(y)}{T^q(y)}\frac{g^\perp}{f_1}, \\
  & \langle \cos\varphi \rangle_{L} = x\kappa_M k_{T M} \frac{\tilde T^q(y)}{T^q(y)}\frac{g^\perp_L}{f_1}, \\
  & \langle \sin\varphi \rangle_{L} =- x\kappa_M k_{T M} \frac{\tilde T^q(y)}{T^q(y)}\frac{f^\perp_L}{f_1}, \\
  & \langle \cos\varphi_S \rangle_{T} =-x\kappa_M \frac{\tilde T^q(y)}{T^q(y)}\frac{g_T}{f_1}, \\
  & \langle \sin\varphi_S \rangle_{T} =-x\kappa_M \frac{\tilde T^q(y)}{T^q(y)}\frac{f_T}{f_1}, \\
  & \langle \cos(2\varphi-\varphi_S) \rangle_{T} = x\kappa_M \frac{k_{T M}^2}{2}\frac{\tilde T^q(y)}{T^q(y)}\frac{g^\perp_T}{f_1}, \\
  & \langle \sin(2\varphi-\varphi_S) \rangle_{T} =- x\kappa_M \frac{k_{T M}^2}{2}\frac{\tilde T^q(y)}{T^q(y)}\frac{f^\perp_T}{f_1}.
\end{align}
We note that the azimuthal asymmetries discussed above are defined for both the electron scattering process and the positron scattering process. For the former,  $U$-type quarks are involved in that process and $D$-type quarks are involved for the latter.


\subsection{The intrinsic asymmetries}

In addition to azimuthal asymmetries, we also define intrinsic asymmetries in the $eN$ collinear frame to explore the imbalance of the transverse momentum of the incident quark in a nucleon. 
Note that the transverse momentum of the incident quark (jet), which lies in the $x–y$ plane, can be decomposed as
\begin{align}
 & k_T^{x}=k_T \cos\varphi, \\
 & k_T^{y}=k_T \sin\varphi.
\end{align}
Therefore, we can define $k_T^{x} (-x)-k_T^{x} (+x)$ to quantify the difference of the transverse momentum between the negative $x$ and positive $x$ directions. The difference in the $y$-direction is defined similarly. To be explicit, we present the general definitions of the intrinsic asymmetries,
\begin{align}
A^x &= \frac{ \int_{-\pi/2}^{\pi/2}d\varphi~ d\tilde{\sigma}-\int_{\pi/2}^{3\pi/2} d\varphi ~d\tilde{\sigma} }{ \int_{-\pi/2}^{\pi/2}d\varphi ~d\tilde{\sigma}_U +\int_{\pi/2}^{3\pi/2} d\varphi~ d\tilde{\sigma}_U}, \label{f:akx}\\
A^y &= \frac{\int_{0}^{\pi} d\varphi~ d\tilde{\sigma}-\int_{\pi}^{2\pi} d\varphi ~d\tilde{\sigma} }{\int_{0}^{\pi} d\varphi ~d\tilde{\sigma}_U +\int_{\pi}^{2\pi} d\varphi~ d\tilde{\sigma}_U}. \label{f:aky}
\end{align}
The $A^x$ and $A^y$ lead to asymmetries in the $x$-direction and $y$-direction, respectively.

According to the definitions,  four kinds of asymmetries are obtained. They are
\begin{align}
 & A_{U}^x = -\frac{4x\kappa_M k_{T M}}{\pi} \frac{\tilde T^q(y)}{T^q(y)}\frac{f^\perp}{f_1}, \label{f:auux} \\
 & A_{U}^y = -\frac{4x\kappa_M k_{T M}}{\pi} \frac{\tilde T^q(y)}{T^q(y)}\frac{g^\perp}{f_1}, \label{f:auuy} \\
 & A_{L}^x =  \frac{4x\kappa_M k_{T M}}{\pi} \frac{\tilde T^q(y)}{T^q(y)}\frac{g_L^\perp}{f_1}, \label{f:aulx} \\
 & A_{L}^y = -\frac{4x\kappa_M k_{T M}}{\pi} \frac{\tilde T^q(y)}{T^q(y)}\frac{f_L^\perp}{f_1}. \label{f:auly} 
\end{align}
We find that the intrinsic asymmetries are all twist-3 measurable quantities.

\subsection{The charge asymmetry}

From Eq. (\ref{f:crosslead}), we can write down the differential cross sections for the electron inelastic scattering and the positron inelastic scattering, respectively. They are
\begin{align}
  d\tilde{\sigma}(e^-)
  = &\frac{\alpha_{\rm em}^2 }{y Q^4} A_W  4x \Big\{ \left[A(y)+C(y)\right]f_1^u \nonumber\\ &+ \left[A(y)-C(y)\right] \left(f_1^{\bar{d}}+ f_1^{\bar{s}}\right) \Big\}, \label{f:crossele} \\ 
  d\tilde{\sigma}(e^+)
  = & \frac{\alpha_{\rm em}^2 }{y Q^4} A_W  4x \Big\{ \left[A(y)-C(y)\right]f_1^{\bar{u}} \nonumber\\ &+ \left[A(y)+C(y)\right]\left(f_1^d + f_1^{s}\right)\Big\}, \label{f:crosspos}
\end{align}
where $d\tilde{\sigma}=d\sigma_{t2}/d\eta d^2l'_T d^2 j_T$, superscripts denote quark flavors and only light flavors are considered here. According to Eqs. (\ref{f:crossele}) and (\ref{f:crosspos}), we introduce the charge asymmetry which is defined as the ratio of the difference to the sum of the differential cross sections of the electron inelastic scattering and the positron inelastic scattering, 
\begin{align}
 A^C=\frac{d\tilde{\sigma}(e^-)-d\tilde{\sigma}(e^+)}{d\tilde{\sigma}(e^-)+d\tilde{\sigma}(e^+)}.
\end{align}
After a simple calculation, we obtain
\begin{align}
 A^C=\frac{\left[A(y)+C(y)\right]F_{N1} +\left[A(y)-C(y)\right]F_{N2}}{\left[A(y)+C(y)\right]F_{D1} +\left[A(y)-C(y)\right]F_{D2}}, \label{f:Rdelta}
\end{align}
where  
\begin{align}
  & F_{N1}=f_1^u-f_1^{\bar{u}}, \label{f:fN1} \\
  & F_{N2}=f_1^{\bar{d}}+f_1^{\bar{s}}- f_1^d - f_1^{s}, \label{f:fN2}\\
  & F_{D1}=f_1^u+f_1^{\bar{u}}, \label{f:fD1} \\
  & F_{D2}=f_1^{\bar{d}}+f_1^{\bar{s}}+ f_1^d + f_1^{s}. \label{f:fD2}
\end{align}
If we define $\delta f^q= f^q-f^{\bar{q}}$ with $q= u, d, s$, the numerator in Eq. (\ref{f:Rdelta}) can be rewritten as
\begin{align}
    {\cal{N}}= \left[A(y)+C(y)\right]\delta f^u - \left[A(y)-C(y)\right]\left(\delta f^d+ \delta f^s\right).
\end{align}
We notice that $A^C$ not only provides a sensitive probe for valence quark distribution funcitons but also reveal violations of strange-antistrange symmetry.

\begin{figure}
\centering \includegraphics[width=0.8\linewidth]{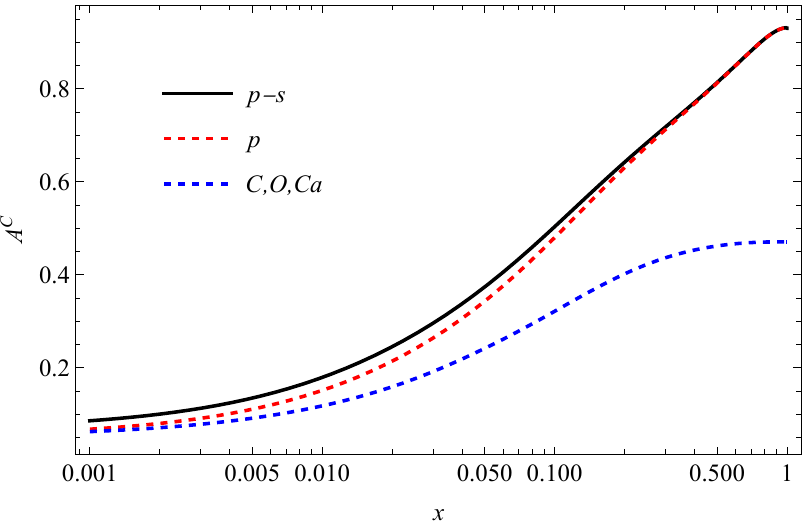}
\caption{Numerical estimations of ratio $A^C$ with respect to $x$. Fraction $y$ and $k_T$ are taken as $y=0.4$ and $k_T=0.5$ GeV. $p-s (p)$ denotes numerical estimates without (with) considering strange and antistrange quarks. $C, O, Ca$ denote estimates for the scattering of nuclei, carbon, oxygen and calcium.}\label{fig:A-x}
\end{figure}

\begin{figure} 
\centering \includegraphics[width=0.8\linewidth]{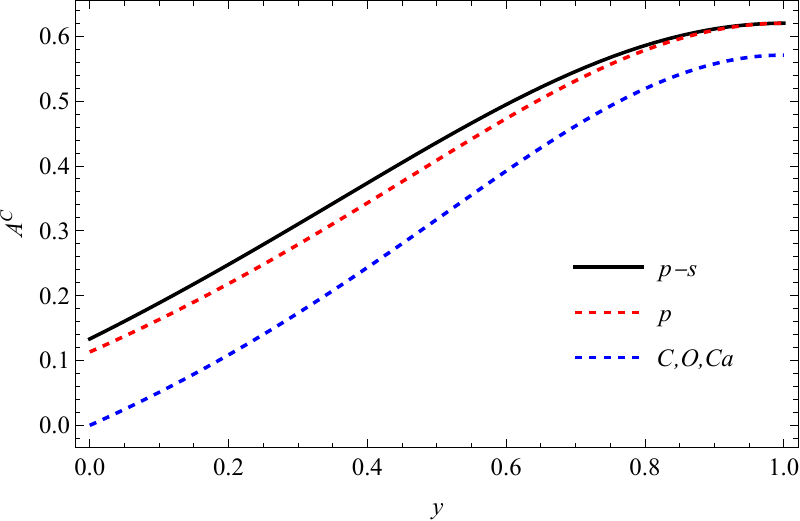}
\caption{Numerical estimations of ratio $A^C$ with respect to $y$. Fraction $x$ and $k_T$ are taken as $x=0.05$ and $k_T=0.5$ GeV. $p-s (p)$ denotes numerical estimates without (with) considering strange and antistrange quarks. $C, O, Ca$ denote estimates for the scattering of nuclei, carbon, oxygen and calcium.}\label{fig:A-y} 
\end{figure}

\begin{figure}
\centering \includegraphics[width=0.8\linewidth]{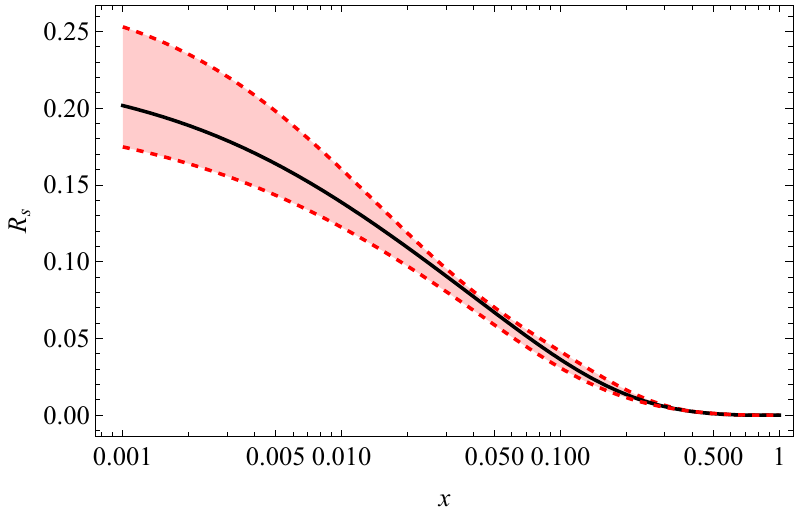}
\caption{Numerical estimations of ratio $R_s$ with respect to $x$.  Fraction $y$ and $k_T$ are taken as $y=0.4$ and $k_T=0.5$ GeV. The band denotes uncertainties caused by PDFs.}\label{fig:Rs} 
\end{figure}

To have an intuitive impression of the charge asymmetry defined above, we present numerical estimates in Fig. \ref{fig:A-x} and Fig. \ref{fig:A-y}. Our estimations are based on the Gaussian ansatz for $f_1(x, k_T)$, i.e.,
\begin{align}
 & f_1(x, k_T) =\frac{1}{\pi {\Delta}^2} f_1(x) e^{-\vec k_T^2/{\Delta}^2}, 
\end{align}
where $f_1(x)$ is taken from CTEQ18 \cite{Hou:2019efy} for proton and from EPPS21  \cite{Hou:2019efy,Eskola:2021nhw}  for carbon, oxygen, and calcium. The average squared transverse momenta are taken as $\Delta_u^2= \Delta_d^2=0.34$ GeV$^2$, $\Delta_{\bar{u}}^2=\Delta_{\bar{d}}^2=0.63$ GeV$^2$, and $\Delta_s^2=\Delta_{\bar{s}}^2=0.22$ GeV$^2$ \cite{Anselmino:2005nn,Signori:2013mda,Anselmino:2013lza,Cammarota:2020qcw,Bacchetta:2022awv,Bacchetta:2024qre} for numerical estimates. In Fig. \ref{fig:A-x} and Fig. \ref{fig:A-y}, black solid lines show numerical estimates without considering strange and antistrange quarks while red dashed lines show estimates with considering strange and antistrange quarks. To see the difference clearly, we introduce the ratio, 
\begin{align}
    R_s=\frac{A^C(s, \bar{s} =0)-A^C(s, \bar{s} \neq 0) }{A^C(s, \bar{s} \neq 0)},
\end{align}
and show it in Fig. \ref{fig:Rs}. We notice that strange and antistrange quarks have significant influence on $A^C$, especially at sea quark region. The band denotes uncertainties caused by 
PDFs.

Blue lines in Fig. \ref{fig:A-x} and Fig. \ref{fig:A-y} show estimates of $A^C$ for inelastic scattering processes of nuclei, carbon, oxygen and calcium. They are isoscalar nuclei with $N=Z$, the proton number equals to neutron number. From these two figure, we see the difference between $A^C_{N\neq Z}$ and $A^C_{N=Z}$ becomes samller and smaller as $x$ becomes smaller.
We also note here that the charge asymmetry defined in this part is independent of the type of target nucleus if is has the same number of neutrons and protons.


\section{Summary} \label{sec:summary}

In this paper, we present a systematic calculation of the charged-current jet-production SIDIS process in the $eN$ collinear frame. The $eN$ collinear frame is defined such that the target travels along the $+z$ direction, while the incoming lepton travels along the $-z$ direction. The scattered (anti-)neutrino lies in the $x-z$ plane which is known as the lepton plane.  The differential cross section is first expressed in terms of structure functions and then expressed in term of TMDs at tree level twist-3. Since the $W$-boson gains the transverse momentum component $q_T$ in the $eN$ collinear frame, the gauge invariance of hadronic tensor becomes difficult to verify directly. To achieve this goal, we present a systematic calculation of how to obtain the gauge invariant hadronic tensor at twist-3 level. 
By comparing these two forms, we obtain a set of relationships between structure functions and the TMDs, shown in Eqs.  (\ref{f:FU})-(\ref{f:FTcos2}). 
We also calculate azimuthal asymmetries and intrinsic asymmetries. Two leading-twist and eight twist-3 azimuthal asymmetries are obtained. However, intrinsic asymmetries are all twist-3 quantities. They provide more measurable quantities for extracting TMDs. 
We also introduce the charge asymmetry $A^C$, which is defined as the ratio of the difference to the sum of the differential cross sections of the electron semi-inclusive deep inelastic scattering and the positron semi-inclusive deep inelastic scattering. We find that $A^C$ can be used to determine the valence quark distribution functions and the violation of strange-antistrange symmetry. Numerical values show that contributions form strange and antistrange quarks become significant as fraction $x$ decreases.

\section*{Acknowledgements}
This work was supported by the National Natural Science Foundation of China (Grant No. 12405103, 12447132) and the Youth Innovation Technology Project of Higher School in Shandong Province (2023KJ146).

\end{document}